\documentclass[reprint, floatfix, twocolumn, amsmath, amssymb, amsfonts, superscriptaddress, pra]{revtex4-2}

\usepackage{graphicx}
\usepackage{comment}

\usepackage{siunitx}
\usepackage{soul}
\usepackage{xcolor}

\usepackage{hyperref}
\hypersetup{
    colorlinks=true,
    linkcolor=blue,
    filecolor=magenta,      
    urlcolor=cyan,
    citecolor=blue,
}

\begin{document}

\title{Scaling native entanglement generation in layered semiconductors with quasi-phase matching}

%
\author{Benjamin Braun}
\affiliation{Vienna Center for Quantum Science and Technology, University of Vienna, Faculty of Physics, 1090 Vienna, Austria}
\affiliation{Vienna Doctoral School in Physics, University of Vienna, Faculty of Physics, 1090 Vienna, Austria}
\author{Andrea Alessandrini}
\altaffiliation{{\it Present address:} { ICFO-Institut de Ci{\`e}ncies Fot{\`o}niques, The Barcelona Institute of Science and Technology, 08860 Castelldefels (Barcelona), Spain}}
\affiliation{Department of Physical and Chemical Sciences, University of L’Aquila, Via Vetoio, 67100 L’Aquila, Italy}

\author{Josip Bajo}
\affiliation{Vienna Center for Quantum Science and Technology, University of Vienna, Faculty of Physics, 1090 Vienna, Austria}
\affiliation{Vienna Doctoral School in Physics, University of Vienna, Faculty of Physics, 1090 Vienna, Austria}

\author{Philipp K. Jenke}
\affiliation{Vienna Center for Quantum Science and Technology, University of Vienna, Faculty of Physics, 1090 Vienna, Austria}
\affiliation{Vienna Doctoral School in Physics, University of Vienna, Faculty of Physics, 1090 Vienna, Austria}
\author{Leone di Mauro Villari}
\affiliation{CNR-SPIN, c/o Dip.to di Scienze Fisiche e Chimiche, Via Vetoio, Coppito (L’Aquila) 67100, Italy}
\author{Birui Yang}
\affiliation{Department of Physics, Columbia University, New York, NY 10027, USA}
\author{Zhi Hao Peng}
\affiliation{Department of Mechanical Engineering, Columbia University, New York, NY 10027, USA}
\author{P. James Schuck}
\affiliation{Department of Mechanical Engineering, Columbia University, New York, NY 10027, USA}
\author{Cory R. Dean}
\affiliation{Department of Physics, Columbia University, New York, NY 10027, USA}
\author{Andrea Marini}
\email{andrea.marini@univaq.it}
\affiliation{CNR-SPIN, c/o Dip.to di Scienze Fisiche e Chimiche, Via Vetoio, Coppito (L’Aquila) 67100, Italy}
\affiliation{Department of Physical and Chemical Sciences, University of L’Aquila, Via Vetoio, 67100 L’Aquila, Italy}
\author{Philip Walther}
\affiliation{Vienna Center for Quantum Science and Technology, University of Vienna, Faculty of Physics, 1090 Vienna, Austria}
\affiliation{Christian Doppler Laboratory for Photonic Quantum Computer \& Research Network Quantum Aspects of Space Time (TURIS), University of Vienna, Faculty of Physics, 1090 Vienna, Austria}
\author{Chiara Trovatello}
\email{chiara.trovatello@polimi.it}
\affiliation{Department of Mechanical Engineering, Columbia University, New York, NY 10027, USA}
\affiliation{Dipartimento di Fisica, Politecnico di Milano, Piazza L. da Vinci 32, I-20133 Milano, Italy}

\author{Lee A. Rozema}
\email{lee.rozema@univie.ac.at}
\affiliation{Vienna Center for Quantum Science and Technology, University of Vienna, Faculty of Physics, 1090 Vienna, Austria}

\begin{abstract}
Efficient generation of entangled photons typically relies on spontaneous parametric down-conversion (SPDC) in phase-matched macroscopic nonlinear media.
However, generating entanglement under phase-matching constraints requires additional bulk optics or interferometers.
In contrast, ultrathin van der Waals semiconductors - such as transition metal dichalcogenides (TMDs) - exhibit strong enough optical nonlinearities for SPDC to be observed from subwavelength-thick media, thereby bypassing conventional phase-matching constraints.
In this microscopic domain, the intrinsic crystal symmetry governs the nonlinear optical response, enabling the native generation of polarization-entangled photon pairs. 
However, generating these states efficiently has been fundamentally restricted by the material's coherence length ($L_c$), which limits the attainable conversion efficiency. 
Here, we investigate periodically-poled TMDs (PPTMDs) designed to scale up this interaction via quasi-phase matching.
We demonstrate that mechanically flipping the sign of the nonlinearity at precise intervals of $L_c$ introduces quasi-phase matching, that scales the pair-production rate while preserving the pristine, symmetry-generated polarization entanglement, with fidelities exceeding 99\%.
Backed by a rigorous theoretical model, our work clarifies the interplay between crystal symmetry and propagation effects in thin nonlinear media, providing a  new avenue for engineering quantum light in nanophotonic systems.
\end{abstract}


\maketitle

\section{Introduction}

From photonic quantum computing to quantum communication and quantum sensing, photonic entanglement lies at the heart of a plethora of different quantum technologies \cite{o2009photonic,wang2025scalable}. Entangled photon pairs (EPPs) for such applications are typically generated with second- ($\chi^{(2)}$) and third-order ($\chi^{(3)}$) spontaneous nonlinear optical processes~\cite{Meyer2020Multiplexing}, such as spontaneous parametric down-conversion (SPDC) and spontaneous four-wave mixing (SFWM), respectively. SPDC - the workhorse for photonic quantum entanglement - is typically carried out in bulk non-centrosymmetric media, such as $\beta$-barium borate (BBO) \cite{Zhong201812photon}, periodically-poled potassium titanyl phosphate (PPKTP) \cite{ greganti2018tuning, Zhao2025Direct} and periodically-poled lithium niobate (PPLN) \cite{Zhao2020ppln}, which require long nonlinear interaction lengths (millimeters to centimeters) \cite{Deng2023Gaussian} to achieve high pair-production rates.
Moreover, these crystals do not natively produce entanglement in polarization, requiring multiple crystals and/or interferometric geometries, as well as compensation optics, to generate entangled states and to correct for spatial and temporal walk-off effects \cite{Kim2006, Kwiat1999, Anwar2021}.
The search for better EPP sources motivates the exploration of novel nonlinear platforms that combine high conversion efficiencies, ultracompact footprints, and native high-fidelity entanglement \cite{Trovatello2024, deAbajo2025}. 

Enabled by the exceptionally strong second-order nonlinearities of layered van der Waals materials, \textit{e.g.}, transition metal dichalcogenides (TMDs) \cite{autere2018nonlinear}, a series of recent experiments has demonstrated the direct generation of high-fidelity polarization-entangled photon pairs from nonlinear media only a few hundreds of nanometers thick \cite{Weissflog2024, Feng2024, Liang2025}.
In this sub-wavelength regime, phase mismatch is negligible and the properties of the generated quantum state are governed primarily by the crystal symmetry.
For example, the experiments in Refs. \cite{Weissflog2024, Feng2024, Liang2025} exploit crystals with C$_{3v}$ symmetry, which directly enables the generation of polarization entanglement \cite{Weissflog2024}.
However, bulk crystals possessing the same symmetry, such as BBO, do not produce polarization entanglement, because phase matching in bulk crystals generally selects only a subset of the nonlinear tensor elements.
At the same time, other sub-wavelength nonlinear materials lacking the required symmetry do not exhibit direct polarization-entangled pair generation \cite{guo2024polarization, joshi2026air}.
Together, these observations highlight a key advantage of ultrathin nonlinear media: the crystal symmetry can directly determine the quantum state, opening new opportunities for photonic quantum-state engineering.

Although these ultrathin materials have great potential, their efficiencies cannot be scaled with increasing thickness beyond the so-called coherence length $L_c$ \cite{boyd}, which for 3R-MoS$_2$, is $L_c \sim \SI{500}{nm}$ at 1520 nm\cite{Xu2022}. In fact, at thicknesses comparable to $L_c$, phase mismatch becomes relevant, and pushing to higher efficiencies requires phase matching the nonlinear interactions.

Recent work has demonstrated quasi-phase-matched generation of photon pairs via SPDC in periodically-poled 3R-TMDs (PPTMDs)~\cite{Trovatello2025}, by stacking multiple slabs of 3R-MoS$_2$ each with a thickness equal to $L_c$ and an inverted sign of the nonlinearity $\hat{\chi}^{(2)}$. 
Preserving entanglement in PPTMDs is challenging, requiring the material to be patterned while preserving its crystal quality, and the alignment between periodically inverted domains to be sufficiently precise to ensure the delicate polarization structure of the two-photon state remains intact \cite{Tang2024,Trovatello2025}.

Here we study the entanglement fidelity and rate enhancement of SPDC in PPTMDs, showing that high-precision slab-to-slab alignment substantially increases the photon-pair generation rate, while preserving the high-fidelity EPP generation - previously only observed for lengths shorter than $L_c$.
We demonstrate polarization-entangled states with fidelities exceeding 99\%, confirming that the high-quality entanglement intrinsic to ultrathin materials is maintained even at increased interaction lengths. 
In addition, we present and validate a comprehensive theoretical model, which quantitatively predicts the generation efficiencies and polarization-state properties.  
Our work demonstrates the viability of periodic-poling in van der Waals materials as a strategy to increase generation efficiencies while preserving their intrinsic polarization-entangled photon pair generation.


\begin{figure}[t]
    \centering
    \includegraphics[width=1\linewidth]{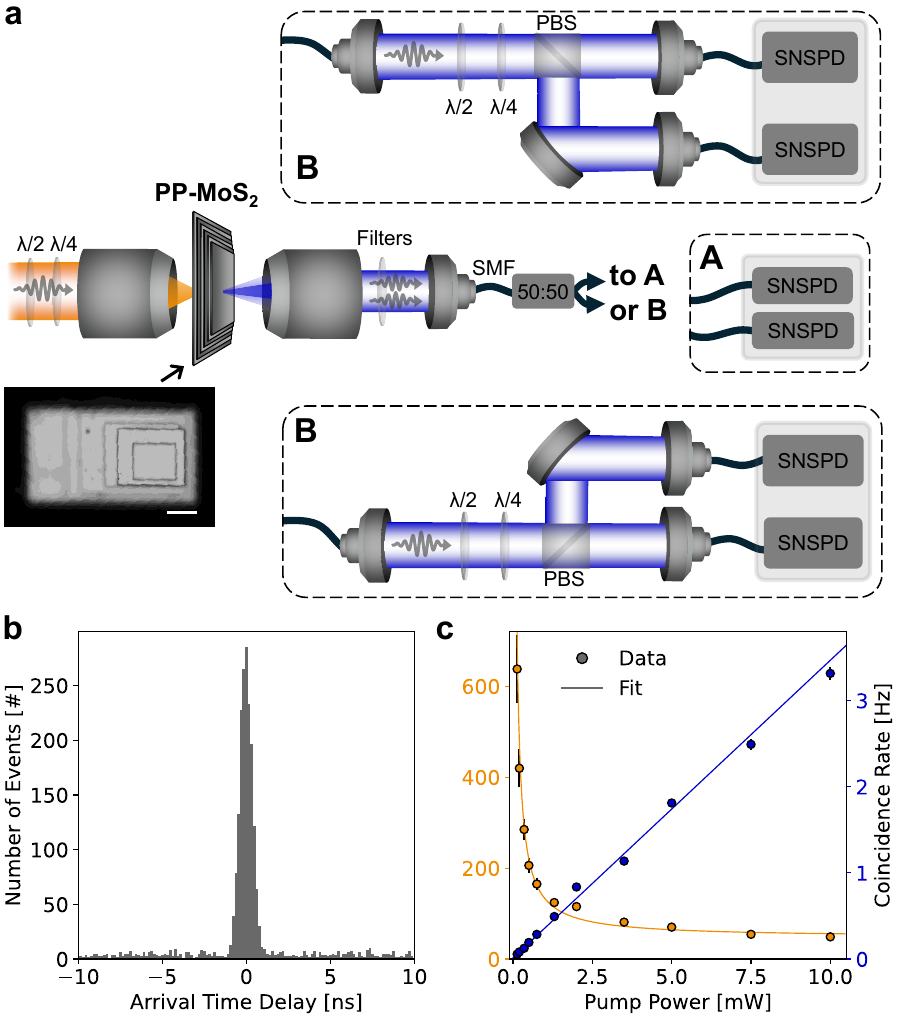}
    \caption{\scriptsize a) Experimental schematic: A continuous wave (CW) pump laser (\SI{780}{nm}) is focused onto the PPTMD (see inset for a micrograph, scalebar \SI{10}{\mu m}) on a SiO$_2$ substrate using an aspheric lens (f=\SI{3.1}{mm}). The emitted signal is collected collinearly and collimated using a microscope objective (NA=0.85). A combination of long-pass (\SI{1150}{nm} cut-off) and band-pass ($1560 \pm \SI{6}{nm}$) filters rejects the pump beam and photoluminescence background. The photon pairs are coupled to single-mode fiber (SMF) and sent to a fiber-based beam-splitter (50:50) which probabilistically separates the pairs into two spatial modes. From here, they can be directly detected (option A) using super-conducting nanowire single-photon detectors (SNSPDs) to confirm the presence of SPDC. Alternatively, the photon pairs can be routed to two polarization tomography arms (option B). The results are shown in Fig. 2a.
    The measurement bases of each arm are set using a half-wave-plate ($\lambda$/2), a quarter-wave-plate ($\lambda$/4) and a polarizing beam-splitter (PBS). 
    b) Arrival time histogram of generated photons in the PPTMD using a pump power of \SI{7}{mW}. The width of the peak comes from timing jitter of the detection system. c) Coincidence Rate and Coincidence-to-Accidental-Ratio (CAR) scaling of the detected signal in dependence of the pump power. Errorbars are calculated from poissonian statistics. Some errorbars are not visible as they are smaller than the markers used. Data reproduced from \cite{Trovatello2025}.
    }
    \label{fig:fig1}
\end{figure}

\begin{figure*}[t]
    \centering
    \includegraphics[width=.8\linewidth]{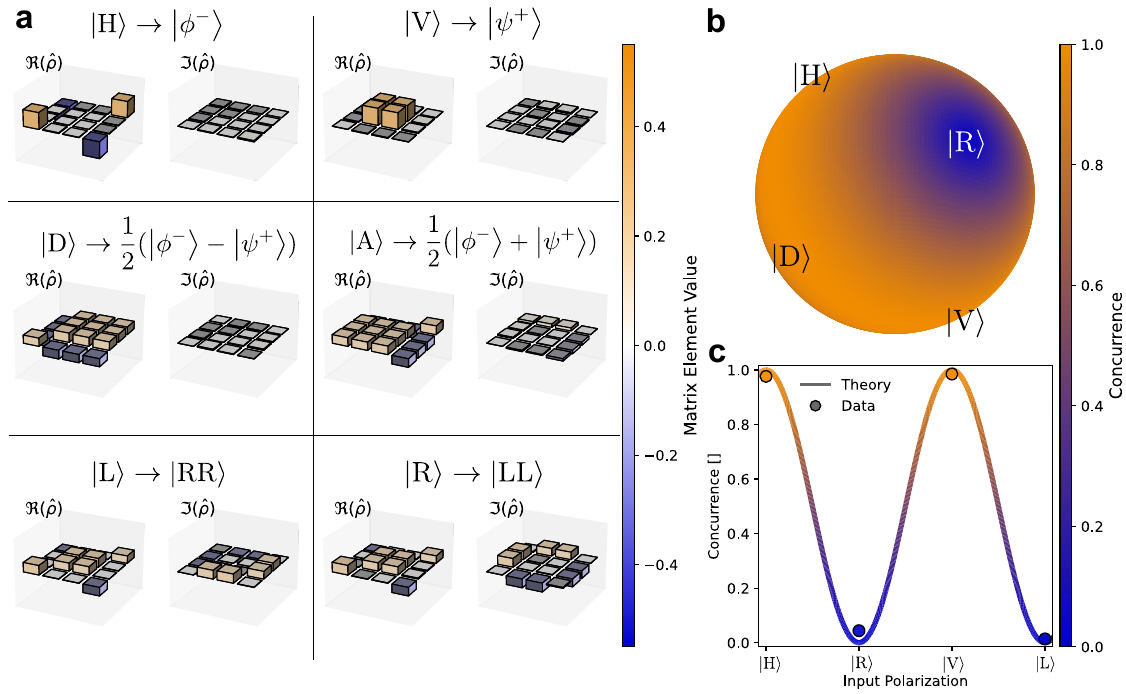}
    \caption{\scriptsize a) Measured density matrices of the 2-photon states generated in PP-MoS$_2$: The measurements shown were carried out on the 6$^\text{th}$ slab using a pump power of \SI{8}{mW} and a band-pass filter ($1560\pm\SI{6}{nm}$). We define the x-axis of the sample as the H-polarization (sample frame). Aligning the pump polarization along this axis results in the maximally entangled Bell-state $\left|\phi^-\right>$. Going to V-polarization in the sample frame generates the $\left|\psi^+\right>$-state. At diagonal polarization, a coherent superposition of these states is generated, which is still maximally entangled. Going to circular polarization completely erases the entanglement, resulting in a fully separable state. b) Simulated concurrence heatmap on the Poincaré sphere of the pump polarization: Moving along the equator of linear pump polarizations generates maximally entangled states in a coherent superposition of $\left|\psi^+\right>$ and $\left|\phi^-\right>$. Moving to elliptical pump polarization reduces the concurrence, at circular pump polarizations the entanglement is entirely erased. c) Expected and measured concurrence values of polarization states generated by different input polarizations moving along the $\left|\text{H}\right>$-$\left|\text{R}\right>$-$\left|\text{V}\right>$-$\left|\text{L}\right>$ equator on the Poincaré sphere. The errorbars of the measured data are numerically estimated via Monte-Carlo simulation performed on the tomography data and do not appear as they are smaller than the markers used.}
    \label{fig:fig2}
\end{figure*}

\section{Polarization Entanglement} 
In this work we use 3R-MoS$_2$, which is a TMD belonging to the aforementioned C$_{3v}$ symmetry group. As such, it exhibits the following non-zero in-plane elements in the 2$^{\text{nd}}$-order nonlinear tensor $\chi^{(2)}_{xxx}=-\chi^{(2)}_{xyy}=-\chi^{(2)}_{yxy}=-\chi^{(2)}_{yyx}$. 
We will now show how this symmetry generates intrinsic polarization entangled photon pairs for any linearly polarized pump photon \cite{Weissflog2024}.
In this notation, the last index refers to the pump polarization, and the first two indices represent the signal and idler polarizations.
This means, that if the pump is polarized along $x$, either $\chi^{(2)}_{xxx}$ can contribute a signal and idler photon both polarized along $x$ or $\chi^{(2)}_{yyx}$ can contribute a signal and idler photon both polarized along $y$. 
In other words, the signal and idler photons' polarization will lie either along $x$ or $y$, which is a polarization $\phi$-Bell state.
If, on the other hand, the pump is polarized along $y$, either $\chi^{(2)}_{xyy}$ or $\chi^{(2)}_{yxy}$ will contribute.
This results in signal and idler photons with opposite polarization, which is a $\psi$-Bell state.

In general, the polarization of the signal and idler fields can be written as:
\begin{eqnarray*}
& \hat{P}_{\text{signal},\alpha} = \epsilon_0 \chi^{(2)}_{\alpha\beta\gamma} E_{\text{pump},\gamma}\hat{E}^{\dagger}_{\text{idler},\beta}, \\
& \hat{P}_{\text{idler},\beta} = \epsilon_0 \chi^{(2)}_{\alpha\beta\gamma} E_{\text{pump},\gamma}\hat{E}^{\dagger}_{\text{signal},\alpha},
\end{eqnarray*}
where ${\rm Re}[{\bf E}_{\text{pump}}({\bf r})e^{-i\omega_{\rm pump}t}]$ is the the pump field with angular frequency $\omega_{\rm pump}$, $[ (1/2)\hat{\bf E}^{\dagger}_{\text{signal,idler}}e^{-i\omega_{\rm signal,idler}t} + {\rm h.c.}]$ are the signal/idler field operators oscillating with angular frequencies $\omega_{\rm signal,idler}$ satisfying $\omega_{\rm signal} + \omega_{\rm idler} =\omega_{\rm pump}$, and $[(1/2)\hat{\bf P}_{\text{signal,idler}}e^{-i\omega_{\text{signal,idler}}t} + {\rm h.c.}]$ are the nonlinearly generated signal/idler polarization operators in the material, respectively. 
Assuming a general polarization for the pumping field 
\begin{equation*}
    {\bf E}_{\text{pump}}=A(\cos\theta\hat{\bf e}_x + \sin\theta e^{i\phi}\hat{\bf e}_y)
\end{equation*}
and swapping to the Bra-Ket notation identifying $\hat{\bf e}_x\hat{=}\left|\text{H}\right>$ and $\hat{\bf e}_y \hat{=}\left|\text{V}\right>$ yields the general two-photon polarization state 
\begin{align*}
    \left|\psi\right> &= \frac{\cos\theta}{\sqrt{2}}(\left|\text{HH}\right>-\left|\text{VV}\right>)
    -\frac{\sin\theta e^{i\phi}}{\sqrt{2}} (\left|\text{HV}\right>+\left|\text{VH}\right>)\\
    &= \cos\theta\left|\phi^-\right>-\sin\theta e^{i\phi}\left|\psi^+\right>.
\end{align*}
One can freely tune the balance of $\left|\psi^+\right>$ and $\left|\phi^-\right>$ and their relative phase just by adjusting the polarization of the pump photons. 

To illustrate the tunability of this source, we plot a heatmap of the concurrence of the two photon state---a measure for entanglement strength ranging from zero to one---on the Poincaré sphere of the input light is shown in Fig. \ref{fig:fig2}b. On the meridian of linear polarization, the down-converted photons are maximally entangled for any setting. This behavior has been previously observed in flakes with thickness less than $L_c$. 
Moving to elliptical polarizations ($\phi\neq 0$) reduces the concurrence. At fully circular polarizations, the entanglement is erased. In Fig. \ref{fig:fig2}c, we plot the concurrence values when moving along the equator of the Poincaré sphere which crosses the $\left|\text{H}\right>$, $\left|\text{V}\right>$, $\left|\text{L}\right>$ and $\left|\text{R}\right>$ polarizations. At fully circular input polarizations the generated states reduce to $\left|\text{L}\right>_{\rm in} \rightarrow \left|\text{RR}\right>_{\rm out}$ and $\left|\text{R}\right>_{\rm in} \rightarrow \left|\text{LL}\right>_{\rm out}$, which are fully separable states.

Notice that all this behavior stems from the crystal symmetry: if the four tensor elements mentioned above have different magnitudes it will, in general, not be possible to generate either maximally entangled or completely separable states.
For example, recent work in ultra-thin NbOI$_2$ (from the C$_2$ group) required a two-crystal geometry to generate entanglement because of the lacking crystal symmetry \cite{joshi2026air}.
At the same time, crystal symmetry on its own is not necessarily sufficient. 
As mentioned above, other crystals, such as BBO also have the same non-zero tensor elements as the 3R-TMDs \cite{boyd2008nonlinear} but do not exhibit this behavior.
This is because standard nonlinear crystals require macroscopic propagation distances and phase matching to observe photon pair production, whereas TMDs have large enough optical nonlinearities to work in the ultrathin regime.
For example, in the case of Type-I phased-matched BBO only a single tensor element of $\hat{\chi}^{(2)}$ can be phase-matched, which necessitates the use two crystals in a so-called sandwich geometry to produce a polarization entangled state \cite{Kwiat1999}.
The situation is in general similar for bulk quasi-phase matched crystals: one can typically only achieve one type of phase matching (Type-0, Type I, or Type-II).
In contrast, in 3R-TMDs Type-0, Type-I and Type-II processes simultaneously contribute, depending on the pump polarization.
Moreover, in bulk crystals, the macroscopic propagation distances introduce distinguishing information, wherein photons generated from different tensor elements gain different spectral or temporal profiles, caused, for example, by birefringent dispersion.
This degrades the entanglement, even if the photons are initially generated.
This can be partially corrected with compensation optics, as is done in Type-II BBO sources \cite{kwiat1995new}.

To maintain this tunability while surpassing the $L_c$ limitation, here we realize PPTMDs from 3R-MoS$_2$, in a geometry where the pump propagates along the optic axis. Note that this is the natural geometry, as it corresponds to setting the pump perpendicular to the crystal surface.
In this configuration, $x$- and $y$-polarized light experience the same refractive index ensuring no deleterious walk-off effects degrade the entanglement as the length is increased.

To create our PPTMDs, we begin by mechanically exfoliating 3R-MoS$_2$ flakes from its bulk counterpart using the standard dry scotch-tape technique. We use commercial bulk 3R-MoS$_2$ (HQ Graphene) grown by chemical vapor transport (for further details on bulk growth and characterization, see Ref. \cite{Xu2022}). Flakes with suitable shape and size are morphologically characterized via atomic force microscopy, and the flake with thickness closest to the coherence length for our target operation wavelength, \textit{i.e.}, $\SI{1560}{nm}$, is selected and patterned via electron beam lithography and reactive ion etching into rectangular slabs. All slabs are subsequently stacked one on top of each other with alternating macroscopic dipole orientation, to flip the sign of the optical nonlinearity $\chi^{(2)}$ at each coherence length and achieve quasi-phase-matched SPDC (for further details on the PPTMD fabrication see Ref. \cite{Trovatello2025}).
For this work, we use a PPTMD with 3 poling periods (6 slabs). Each slab has thickness of $\sim\SI{570}{nm}$, therefore the 6$^{\text{th}}$ stacking region has total thickness of $\sim\SI{3.4}{\micro m}$. The inset of Fig. \ref{fig:fig1}a shows a micrograph of the sample.

To characterize SPDC from our PPTMD, we use the setup illustrated in Fig. \ref{fig:fig1}a. The PPTMD is pumped by a CW laser at $\SI{780}{nm}$ which is focused onto the sample with an aspheric lens ($f=\SI{3.1}{mm}$). The emitted SPDC is collected with a high numerical aperture (NA) microscope objective ($\text{NA}=0.85$), spectrally filtered using a combination of hard-coated long-pass and band-pass filters, and coupled to a single-mode fiber. A balanced fiber-based beam-splitter probabilistically splits the photon pairs into two spatial modes. From here, photons can either be detected directly in order to measure their timing correlations or each photon can be routed to a separate polarization tomography stage.
In both cases, super-conducting nanowire single-photon detectors (SNSPDs) are used to detect the photons ($\sim \SI{85}{\percent}$ efficiency).

In the first configuration, where we directly detect the two photons after the beam-splitter, we observe the expected narrow coincidence peak at zero time delay in the arrival time histogram, as shown in Fig. \ref{fig:fig1}b. Its width is caused by timing jitter of the detectors and the time-tagger. Events in bins with non-zero time delay correspond to accidental two-photon events between either uncorrelated SPDC photons or background photoluminescence. 
The ratio of the number of coincidences in the zero time delay bin and the average of all other bins is the coincidence-to-accidental ratio (CAR). 
The difference between zero delay bin and the non-zero delay bins is the number of events caused by photon pairs corrected for accidental events, \textit{i.e.}, the real coincidence rate.
All data presented in the following are corrected for accidental coincidence events. 
The CAR exhibits an inverse scaling with the power of the pumping field (orange), while the coincidence rate is linearly dependent on pump power (blue), as expected (see Fig. \ref{fig:fig1}c). 
Both CAR and coincidence rate are strongly affected by the spectral filters used to isolate the SPDC signal. 
The data shown in Fig.~\ref{fig:fig1}c were recorded with a band-pass filter centered at $\SI{1560}{nm}$ with a full-width at half-maximum (FWHM) of $\SI{12}{nm}$. 
By using this filter, we reduce the brightness of our spectrally broad photon-pair source - which emits photon pairs over a spectral width of at least \SI{400}{nm} \cite{Trovatello2025} - but we also remove unwanted photoluminescence background, effectively increasing the CAR. This background is peaked around $\SI{850}{nm}$, and exponentially decays into telecom and infrared wavelengths \cite{Weissflog2024, Feng2024}.

\begin{table}
\begin{tabular}{||c | c | c | c||} 
 \hline
 Input Pol. & Purity & Concurrence & Fidelity\\
 \hline\hline
 $\left|\text{H}\right>$ & $0.989\pm0.005$ & $0.985\pm0.005$ & $0.994\pm0.001$\\ 
 \hline
 $\left|\text{V}\right>$ & $0.978\pm0.006$ & $0.976\pm0.004$ & $0.993\pm0.001$\\
 \hline
 $\left|\text{D}\right>$ & $0.985\pm0.005$ & $0.979\pm0.005$ & $0.994\pm0.001$\\
 \hline
 $\left|\text{A}\right>$ & $0.992\pm0.005$ & $0.961\pm0.005$ & $0.985\pm0.002$\\
 \hline
 $\left|\text{L}\right>$ & $0.978\pm0.004$ & $0.003\pm0.014$ & $0.993\pm0.001$\\
 \hline
 $\left|\text{R}\right>$ & $0.975\pm0.005$ & $0.044\pm0.023$ & $0.992\pm0.001$\\
 \hline
\end{tabular}
\caption{\scriptsize Purity, Concurrence and Fidelity (with regard to the target state) values of the 2-photon polarization states, generated from different pump polarization settings. They correspond to the density matrices shown in Fig. \ref{fig:fig2}a. The error margins are numerically estimated via Monte-Carlo simulation performed on the tomography data.}
\label{fig:fig3}
\end{table}

We verify that our poling preserves entanglement by sending the two photons to polarization tomography setups, each consisting of a half-wave plate, a quarter-wave plate and a polarizing beam-splitter (Fig. \ref{fig:fig1}a)~\cite{JamesMeasurement2001}. 
For these measurements, we use the same $\SI{1560}{nm} \pm \SI{12}{nm}$ filter as above.
To acquire our quantum state tomography data, we use a narrow coincidence window of \SI{647}{ps}---including the coincidence peak but rejecting accidental events---and record the total number of coincidence counts in all 9 combinations of the required bases. 
For each basis setting we set the integration time to 1200 seconds, corresponding to an approximate total number of counts of 800 per setting. We then use maximum likelihood estimation to evaluate the full density matrix.

We set the pump polarization to all six basis states with regards to the crystalline axis of the material, and perform quantum state tomography on the resulting down-converted photon pairs.
As shown in Fig. \ref{fig:fig2}a, we verify the generation of the expected Bell states $\left|\phi^-\right>=\frac{1}{\sqrt{2}}(\left|\text{HH}\right>-\left|\text{VV}\right>)$, $\left|\psi^+\right>=\frac{1}{\sqrt{2}}(\left|\text{HV}\right>+\left|\text{VH}\right>)$, their balanced superpositions $\frac{1}{\sqrt{2}}(\left|\phi^-\right>\pm\left|\psi^+\right>)$ and the separable states $\frac{1}{\sqrt{2}}(\left|\phi^-\right>\pm i\left|\psi^+\right>)$, in agreement with previous results \cite{Weissflog2024, Feng2024}, but now for a media thickness significantly exceeding $L_c$. 
We observe state fidelities of 0.99 and purities of 0.98 for all states when correcting for accidental coincidence events and unbalanced detector efficiencies (the exact values for all input pump polarizations are reported in Table \ref{fig:fig3}). 
These pristine states are achieved because they stem directly from the $\hat\chi^{(2)}$ of the material with a minimal number of optical elements degrading them or relying on alignment dependent optical setups to generate them altogether.

\begin{figure}[t]
    \centering
    \includegraphics[width=1\linewidth]{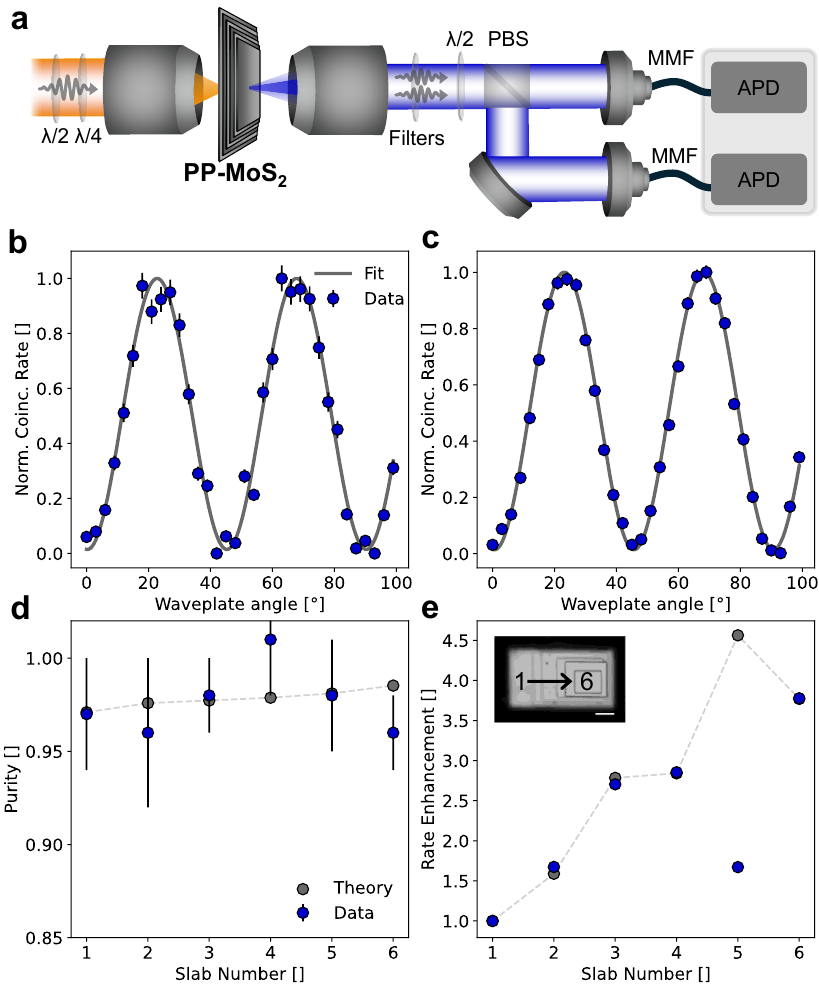}
    \caption{\scriptsize a) Adapted experimental setup: the polarization of the emitted photon pairs is controlled with a half wave-plate ($\lambda$/2), then the photons are sent to a polarizing beam-splitter (PBS). Depending on their polarization state, they may split into the two output modes of the PBS or bunch in one mode. We couple both modes to a multi-mode fiber (MMF, \SI{50}{\mu m} core size) and detect them with In-Ga-As avalanche photo-diodes (APDs). b) \& c) The coincidence rate is measured as a function of the wave-plate angle. We fit the data points to $\cos^{2}(4x)$ function with a phase and background offset, where $x$ is the wave-plate angle, in order to extract the visibility. We show example plots measured on slab 1 (b) and slab 6 (c). The errorbars are calculated from poissonian statistcs, some of them are not visible as they are smaller than the markers used. d) Predicted and measured purity of the generated polarization states in dependence of the slab number. We extract the measured purity from the corresponding visibilities according to equation (\ref{eq_purity}). The errorbars are calculated from the error of the fitting procedure and error propagation. e) Predicted and measured pair generation rate enhancement in dependence of the slab number. See inset for micrograph of the stacked up slabs, scale bar \SI{10}{\mu m}. Errorbars of the measured data are calculated from poissonian statistics. However, they are not visible as they are smaller than the markers used. The dashed lines in d) and e) serve as a guide for the eye and convey no physical meaning.}
    \label{fig:fig4}
\end{figure}

\section{Poling Dependence}
To theoretically study this system, we create a model based on canonical quantization to predict the generation efficiencies and polarization properties of the EPPs.
In the model we neglect both absorption and dispersion of the down-converted photons and perform the quantization of a monochromatic field $\hat{\bf E}_{\rm Q}({\bf r},t)$.
We consider a classical monochromatic pump field ${\bf E}_{\rm P}({\bf r},t)$ in the undepleted pump approximation, at normal incidence, consistent with the experimentally used CW pump beam.
For a full description of our model we refer to the Appendix (section \ref{app_theory}).
Our model enables the design of the photon sources to specific requirements before fabrication.

To compare the model to our experimental data and directly analyze how the periodic-poling affects the properties of the SPDC emission, we use a simplified setup shown in Fig. \ref{fig:fig4}a to directly estimate the fidelity at different points on the sample without performing full quantum state tomography. 
To do so, we use a half-wave plate and a PBS to split our photon-pairs and couple each output port of the PBS to fiber.
As we show in the Appendix (section \ref{app_purity}), assuming that the pump polarization is set to linear, the visibility of the resulting two-photon fringes as the half-wave plate is rotated can be used to directly estimate the purity of the two-photon state.
In particular, we can express the purity $P$ of the two-photon state as a function of the visibility $V$:
\begin{equation}
    P=\frac{1}{2}(1+V^2). \label{eq_purity}
\end{equation}

Intuitively, this relation can be understood as follows. 
If the sample is pumped with, for example, vertical polarization, the Bell state $\left|\psi^{+}\right>$ is generated in a single spatial mode. 
Since this state is anti-correlated in polarization, the photons will always split on the PBS. 
Putting the half-wave plate to the horizontal setting will not affect the polarization of the photons in this case. 
However, rotating the half-wave plate to \ang{22.5}, transforms the $\left|\psi^{+}\right>$ state to $\left|\phi^{-}\right>$. 
In this case, the photon pairs always bunch on the PBS and no coincidence events between the two detectors will occur.
But this requires two-photon interference and entanglement.
If the initial state is not $\left|\psi^{+}\right>$, but in some mixed state the perfect cancellation of coincidences at \ang{22.5} will not occur.
This thus provides a relatively quick method to estimate the entangled state quality, without acquiring a full tomography dataset.

An additional benefit of our simplified measurement apparatus is that it uses free-space polarization optics directly after the sample. Thus, we can couple the photons directly into multi-mode fiber which increases the effective numerical aperture.
In addition to increasing the count rates, the higher numerical aperture allows us to characterize the emission properties over a larger emission cone compared to single-mode fibers. 
Note that, this also makes it necessary to use multi-mode fiber compatible InGaAs avalanche photo-diodes ($\sim \SI{20}{\percent}$ efficiency) rather than the single-mode SNSPDs used for our previous measurements. 
Addressing each slab of the PPTMD source individually, we collect the SPDC efficiency enhancement (Fig. \ref{fig:fig4}e), and the purity of the polarization states (Fig. \ref{fig:fig4}d) as a function of the number of slabs and compare them to the prediction from our simulation.

We then measure the fringe visibility and purity from every slab.  Representative measurements are plotted in Fig. \ref{fig:fig4}b and \ref{fig:fig4}c for slabs one and six, respectively. 
The interference fringes show high visibilities, confirming that the poling does not introduce any measurable decrease to the entanglement quality. 
To quantify this, we plot the purities extracted by fitting to these data versus slab number in Fig. \ref{fig:fig4}d.
Within experimental error, the purity remains constant.
At the same time, Fig. \ref{fig:fig4}e, shows the enhanced two-photon emission rate.
Overall, we find a strong agreement between our theory and experiment, with the exception of a single data point, which we attribute to experimental inaccuracies in addressing specific slabs which have small lateral size, \textit{i.e.}, comparable to the laser spotsize (see inset of Fig. \ref{fig:fig1}a). 

\section{Discussion}We have demonstrated the direct generation of Bell-states $\left|\psi^+\right>$ and $\left|\phi^-\right>$ as well as fully separable states $\frac{1}{2}\left|\psi^+ \pm i\phi^-\right>$ and their superpositions at the relevant telecom wavelength and with remarkably high fidelities in PPTMDs of 3R-MoS$_2$.
We stress that the generation of separable photon pairs is actually non-trivial, as it requires coherent two-photon interference between quantum states emitted from a superposition of tensor elements.
Our results are achieved in a straightforward collinear geometry without the need for any compensating crystal or cumbersome interferometric techniques. 
We have shown that our quasi-phase matching fabrication technique can achieve higher generation efficiencies while maintaining the high purity of the microscopically generated entanglement. Furthermore, we have developed a comprehensive simulation package to  predict generation efficiencies and polarization-state properties, validated by the acquired experimental data.

We observe a fourfold increase in the generation efficiency here, however we have previously measured an increase of more than an order of magnitude \cite{Trovatello2025} in another PPTMD sample with a different thickness. This is due to the strong etalon-like resonances caused by internal reflections within the material and is also predicted by our simulation, without an appreciable drop in the fidelity.
While coincidence rates are still modest, rapid advances in fabrication techniques and yet unexplored possibilities such as cavity enhancements and coupling to internal resonances of the semiconductors offer promising outlooks for the future. 
Coupling to nonlocal metasurfaces also offer an interesting route to enhance the effective nonlinearity by local field enhancement and mode matching \cite{Verre2019, Peng2025}. 
Moreover, van der Waals materials are quite versatile, have already been used to demonstrate nonlinear conversion in waveguiding geometry \cite{Xu2022}, with the possibility to in- and out-couple light using grating couplers \cite{MooshammerXu2024}.
Especially in the context of on-chip integration, which is readily available for van der Waals materials \cite{Azimi2025, Liu2019}, bypassing coupling losses may out scale ordinary SPDC sources for multi-photon applications.


\section*{ACKNOWLEDGEMENTS}

\noindent The authors thank Marco Liscidini, Alessia Stefano, Simone Paganelli for useful discussions and Michael Antesberger for laboratory resources.

This project has received funding from the European Union(HORIZON Europe Research and Innovation Programme,
EPIQUE, No 101135288). Views and opinions expressed are however those of the author(s) only and do not necessarily reflect those of the European Union or the European Commission-EU. Neither the European Union nor the granting authority can be held responsible for them. This  research was also funded in whole or in part by the Austrian Science Fund (FWF)[10.55776/COE1] (Quantum Science Austria), [10.55776/F71] (BeyondC) and [10.55776/FG5] (Research Group 5). For open access purposes, the author has applied a CC BY public copyright license to any author accepted manuscript version arising from this submission. C.T. acknowledges the European Union’s Horizon Europe research and innovation programme under the Marie Skłodowska-Curie PIONEER HORIZON-MSCA-2021-PF-GF grant agreement No 101066108; the Optica Foundation for supporting this research through the 2024 Optica Foundation Challenge Award (q-POLIS); the European Union—NextGenerationEU under the National Quantum Science and Technology Institute (NQSTI) Grant No. PE00000023-q-ANTHEM-CUP H43C22000870001. P.W. acknowledges financial support from the Air Force Office of Scientific Research under award number FA9550-21-1-0355 (Q-Trust) and FA8655-23-1-7063 (TIQI), as well as the Austrian Federal Ministry of Labour and Economy, the National Foundation for Research, Technology and Development and the Christian Doppler Research Association. B.B., J.B. and P.K.J. acknowledges support from the University of Vienna via the Vienna Doctoral School. L.D.M.V. acknowledges funding from the Italian Ministry of University and Research, \emph{Young Researchers program} PLASMXUV Grant No. SOE20240000030- CUP E17G24000510001.

\bibliographystyle{naturemag}
\bibliography{bib} 

\section*{Appendix}
\subsection{Experimental Methods}\label{app_methods}

A CW laser at a wavelength of $\SI{780}{nm}$ (TOPTICA TA pro) is controlled in power and polarization using a a neutral density filter wheel, a half wave-plate and a quarter wave-plate. It is focussed by an aspheric lens with a focal length of \SI{3.1}{mm} (Thorlabs C330TMD-B), resulting in a focal spot with a diameter of $\SI{1.8}{\mu m}$. For the typically used laser power of $\SI{8}{mW}$, this corresponds to an intensity of $\SI{3.14}{GW/m^2}$. The PPTMD (SiO$_2$ substrate facing the focusing lens) is positioned in the focus of the laser beam. The emitted light is collected and collimated by an objective with a numerical aperture of 0.85 (Evident LCPLN100XIR). The residual laser beam as well as photoluminescence emission are rejected with a set of hard-coated filters (Thorlabs FELH1150, Thorlabs FBH1560-12). The remaining signal is coupled to single mode fiber (Corning SMF-28) and sent to a balanced fiber-based beam-splitter (Thorlabs TW1550R5A1). From here, the signal can either directly be routed to superconducting nanowire detectors ($\sim$\SI{85}{\%} detection efficiency at $\SI{1550}{nm}$, photon spot) or routed to two tomography arms as shown in Fig. \ref{fig:fig1}a. In the second measurement setup, the signal is coupled to multi-mode fibers (Thorlabs FG050LGA) and detected by InGaAs avalanche photo diodes ($\sim\SI{20}{\%}$ detection efficiency at \SI{1550}{nm}, ID Quantique ID 220). Click events are read out using a time-tagger (UQDevices Logic16).

\subsection{Details on the Theory Model}\label{app_theory}

In this appendix, we give some details on the theory model employed to account for the experimental results. As we mentioned in the main text, we neglect both absorption and dispersion in correspondence of the down-converted frequencies and perform the quantization of a monochromatic field $\hat{\bf E}_{\rm Q}({\bf r},t)$ \cite{Mandel1995} in the given medium, obtaining
\begin{equation}
    \begin{aligned}
        \hat{\bf E}_{\rm Q}\left({\bf r},t\right)&=i\int d{\bf k}_{\perp}\sum_{f,s=\pm 1}\sqrt{\frac{\hbar \omega^{\rm (Q)}}{2\varepsilon_0}}\times\\
        &\hspace{1.7cm}\times\hat{a}_{{\bf k}_{\perp},f,s}{\bf u}_{{\bf k}_{\perp},f,s}e^{-i\omega^{\rm (Q)} t}+{\rm h.c.},\\
    \end{aligned}
    \label{EMQuantumFieldVac}
\end{equation}
with 
\begin{equation}
    {\bf u}_{{\bf k}_{\perp},f,s}=\hat{\bf n}_{{\bf k}_{\perp},f,s}e^{i{\bf k}_{\perp}\cdot{\bf r}_{\perp}}\frac{e^{fi\sqrt{\frac{\left(\omega^{\rm (Q)}\right)^2}{c^2}\varepsilon_{\rm L}'-k_{\perp}^2}z}}{2\pi\sqrt{\varepsilon_{\rm L}'d_{\rm NL}}},
\end{equation}
where the index ${\rm Q}$ stands for signal or idler field,  ${\bf k}_\perp=(k_x,k_y)$ represents the perpendicular momentum to the $c-$axis which is conserved imposing lateral translational invariance, $f$ stands for forward ($f=1$) or backward ($f=-1$) propagating quantum modes, $s=\pm 1$ identify two orthogonal polarizations, $\omega^{\rm(Q)}$ is the down-converted frequency in the degenerate scenario, $\varepsilon_0$ is the vacuum permittivity, $c$ is the speed of light in vacuum, $\hbar$ is the Plank constant, $\hat{a}_{{\bf k}_{\perp},f,s},\,\hat{a}^{\dagger}_{{\bf k}_{\perp}',f',s'}$ are destruction and creation operators associated to the different bosonic modes satisfying
\begin{equation}
    \begin{aligned}
        &\left[\hat{a}_{{\bf k}_{\perp},f,s},\hat{a}^{\dagger}_{{\bf k}_{\perp}',f',s'}\right]=\delta\left({\bf k}_{\perp}-{\bf k}_{\perp}'\right)\delta_{ff'}\delta_{ss'},\\
        &\left[\hat{a}_{{\bf k}_{\perp},f,s},\hat{a}_{{\bf k}_{\perp}',f',s'}\right]=\left[\hat{a}^{\dagger}_{{\bf k}_{\perp},f,s},\hat{a}^{\dagger}_{{\bf k}_{\perp}',f',s'}\right]=0,
    \end{aligned}
    \label{commRel}
\end{equation}
${\bf u}_{{\bf k}_{\perp},f,s}$ are spatially normalized quantum modes satisfying the inner product
\begin{equation}
\begin{aligned}
    &\langle{\bf u}_{{\bf k}_{\perp},f,s}|{\bf u}_{{\bf k}_{\perp}',f',s'}\rangle=\\
    &\hspace{1cm}=\int d{\bf r}_{\perp}\int_{-d_{\rm NL}/2}^{d_{\rm NL}/2}dz\varepsilon_{\rm L}'{\bf u}_{{\bf k}_{\perp},f,s}^*\left({\bf r}\right){\bf u}_{{\bf k}_{\perp}',f',s'}\left({\bf r}\right)=\\
    &\hspace{1cm}=\delta\left({\bf k}_{\perp}-{\bf k}_{\perp}'\right)\delta_{ff'}\delta_{ss'},
    \end{aligned}
    \label{innProd}
\end{equation}
$\hat{\bf n}_{{\bf k}_{\perp},f,s}$ are unit vectors in a given polarization basis, $\varepsilon_{\rm L}'$ is the real part of the linear response of the 3R-MoS$_2$ at the down-conversion wavelength, and $d_{\rm NL}$ represents the thickness of the nonlinear system. We emphasize that, by fixing the commutation relations in Eq. \eqref{commRel}, then we automatically enforce $\delta_{ff'}$ also in the inner product in Eq. \eqref{innProd}. By strictly imposing this orthogonality over the finite length $d_{\rm NL}$, we are effectively applying a spatial equivalent of the rotating wave approximation, and thus we project the rapidly oscillating spatial cross-terms out of the signal/idler Hilbert space.

 We consider a classical monochromatic impinging field ${\bf E}_{\rm P}({\bf r},t)$ in the undepleted pump approximation, at normal incidence 
\begin{equation}
\begin{aligned}
    {\bf E}\left(z,\omega^{\rm (P)}\right)=&
     \Theta\left(-z-d_{\rm NL}/2\right){\bf E}_{\rm VAC,<}\left(z,\omega^{\rm (P)}\right)+\\
     &+\Theta_{\rm NL}{\bf E}_{\rm NL}\left(z,\omega^{\rm (P)}\right)+\\
     &+\Theta\left(z-d_{\rm NL}/2\right){\bf E}_{\rm VAC,>}\left(z,\omega^{\rm (P)}\right),
\end{aligned}
    \label{LinearClassicalField}
\end{equation}
with
\begin{widetext}
\begin{equation}
\begin{aligned}
    {\bf E}_{\rm VAC,<}=&E_{\rm VAC}^{\rm IN}e^{ik^{\rm (P)}\left(z+d_{\rm NL}/2\right)}\left(\alpha_{\rm IN}\hat{\bf e}_x+\beta_{\rm IN}\hat{\bf e}_y\right)+E_{\rm VAC}^{\rm R}e^{-ik^{\rm (P)} \left(z+d_{\rm NL}/2\right)}\left(\alpha_{\rm R}\hat{\bf e}_x+\beta_{\rm R}\hat{\bf e}_y
        \right),\\
{\bf E}_{\rm NL}=&
\left(E_{\rm NL}^{+,x}\hat{\bf e}_x+E_{\rm NL}^{+,y}\hat{\bf e}_y\right)e^{ik^{\rm (P)}\sqrt{\varepsilon_{\rm L}\left(\omega^{\rm (P)}\right)}\left(z+d_{\rm NL}/2\right)}+\left(E_{\rm NL}^{-,x}\hat{\bf e}_x+E_{\rm NL}^{-,y}\hat{\bf e}_y\right)e^{-ik^{\rm (P)}\sqrt{\varepsilon_{\rm L}\left(\omega^{\rm (P)}\right)}\left(z+d_{\rm NL}/2\right)},\\
{\bf E}_{\rm VAC,>}=&
E_{\rm VAC}^{\rm T}e^{ik^{\rm (P)} \left(z-d_{\rm NL}/2\right)}\left(\alpha_{\rm T}\hat{\bf e}_x+\beta_{\rm T}\hat{\bf e}_y\right),
\end{aligned}
\label{field2}
\end{equation}
\end{widetext}
where ${\bf E}\left(z,t\right)={\rm Re}\left\{{\bf E}\left(z,\omega^{\rm (P)}\right)e^{-i\omega^{\rm (P)}t}\right\}\equiv{\bf E}_{\rm P}\left({\bf r},t\right)$, $k^{\rm (P)}=\omega^{\rm (P)}/c=2\omega^{\rm (Q)}/c$, ${\bf E}_{\rm VAC,<}$ and ${\bf E}_{\rm VAC,>}$ are the vacuum fields before and after the sample, respectively. Moreover, $\alpha_{\rm IN,R,T}$ and $\beta_{\rm IN,R,T}$ are complex constants accounting for mode normalization and for the arbitrary polarization state. They are respectively related to the incident, reflected, and transmitted pump fields. The Heaviside step-function  $\Theta_{\rm NL}=\Theta\left(z+d_{\rm NL}/2\right)-\Theta\left(z-d_{\rm NL}/2\right)$ localizes $\mathbf E_{\rm NL}$ inside the nonlinear crystal, $\varepsilon_{\rm L}\left(\omega^{\rm (P)}\right)$ is the total linear dielectric function at the pump frequency, $E_{\rm VAC}^{\rm IN/R/T}$ are vacuum amplitudes related to the incident, reflected, and transmitted pump fields, $E_{\rm NL}^{\pm,x/y}$ are cartesian components of either forward or backward medium field's amplitudes, and $\hat{\bf e}_{x,y,z}$ are unit vectors along the corresponding cartesian directions. 

To accurately describe the experimental results, the modeling of the periodic poling is of utmost importance. As discussed, the periodic poling is achieved through a mechanical rotation of a $\pi$ angle about the $c-$axis of the stacked slabs. This results in a periodic rotation of the geometrical properties of the TMD. Neglecting the in-plane linear anisotropy, we only have a $\pi$ rotation of the $\hat{\chi}^{(2)}$ tensor
\begin{equation}
    \chi_{{\rm R},i_1i_2i_3}^{(2)}=R_{i_1j_1}R_{i_2j_2}R_{i_3j_3}\chi^{(2)}_{j_1j_2j_3},
    \label{rotation}
\end{equation}
where $R_{ij}$ is the 3D rotation matrix at angle $\pi$ about the $c-$axis. To model SPDC in periodicallay-poled micro-crystals, we consider the coherent superposition of the pump (${\bf E}_{\rm P}$) and the quantum field [$\hat{\bf E}_{\rm Q}({\bf r},t)$], yielding $\hat{\bf E}_{\rm TOT}({\bf r},t)={\bf E}_{\rm P}({\bf r},t)+\hat{\bf E}_{\rm Q}({\bf r},t)$, where the latter embeds both signal and idler contributions within the NL medium. Moreover, assuming a real non-dispersive nonlinear tensor (valid far from interband transitions energies), we can express the second-order nonlinear polarization through the following expression
\begin{equation}
    \begin{aligned}
        \hat{\bf P}^{(2)}({\bf r},t)&=\varepsilon_0\left\{\sum_{n=1}^N\Theta_n(z)\hat{\chi}_n^{(2)}:\hat{\bf E}_{\rm TOT}\otimes\hat{\bf E}_{\rm TOT}\right\},
    \end{aligned}
\end{equation}
where $N$ is the total number of periodically-poled slabs, $\hat{\chi}_n^{(2)}=\hat{\chi}^{(2)}$ if $n$ is odd, $\hat{\chi}_n^{(2)}=\hat{\chi}_{\rm R}^{(2)}$ if $n$ is even, $\Theta_n(z)=\Theta\left[z-d_{\rm NL}\left(-\frac{1}{2}+\frac{n-1}{N}\right)\right]-\Theta\left[z-d_{\rm NL}\left(-\frac{1}{2}+\frac{n}{N}\right)\right]$ is the Heaviside function related to the $n$th slab of the stack, and $d_{\rm NL}=Nd_{\rm TMD}$ with $d_{\rm TMD}$ being the thickness of a single slab. In other words, in order to solve the nonlinear problem, we use the solution to the linear quantum Maxwell's equations [Eq. \eqref{EMQuantumFieldVac}]  and then introduce the effects of the second order nonlinearity perturbatively \cite{Mandel1995}. Moreover, by virtue of the correspondence principle of quantum mechanics, we will consider only the contribution to the nonlinear polarization related to the difference frequency generation process. 
Starting from $\hat{H}_{\rm INT}=(1/2)\int_{\rm TMD}d^3 r\hat{\bf P}^{(2)}({\bf r},t)\cdot \hat{\bf E}_{\rm TOT}({\bf r},t)$ for the interaction Hamiltonian, we eventually retrieve
\begin{equation}
    \begin{aligned}
        \hat{H}_{\rm INT}&=\hbar\int d{\bf k}_{\perp}\int d{\bf k}_{\perp}'\sum_{s,s'=\pm 1}\left[G\left({\bf k}_{\perp},{\bf k}_{\perp}'\right)_{s,s'}\times\right.\\
        &\hspace{1.5cm}\left.\times\left(\hat{a}_{{\bf k}_{\perp},s}^{\rm VAC}\right)^{\dagger}\left(\hat{a}_{{\bf k}_{\perp}',s'}^{\rm VAC}\right)^{\dagger}+{\rm h.c.}\right],
    \end{aligned}
\end{equation}
where 
\begin{widetext}
    \begin{equation}
        \begin{aligned}
            G \left({\bf k}_{\perp},{\bf k}_{\perp}'\right)_{s,s'}&=-\delta(k_y+k_y')\delta\left(k_x+k_x'\right)\frac{\omega^{\rm (Q)}}{8\varepsilon_{\rm L}'}\sum_{n=1}^N\sum_{pqr}\left(\chi_{n,qrp}^{(2)}+\chi_{n,rqp}^{(2)}\right)\sum_{\sigma,\sigma'=\pm 1}\frac{\left(\hat{\bf n}_{{\bf k}_{\perp},1,\sigma}\right)_{q}^*\left(\hat{\bf n}_{{\bf k}_{\perp}',1,\sigma'}\right)_{r}^*}{N}\times\\
        &\times\left(M_{s}^{\sigma}\right)^*\left(M_{s'}^{\sigma'}\right)^*\left\{
            \begin{aligned}
            &E_{{\rm NL}}^{+,p}e^{ik^{\rm (P)}\sqrt{\varepsilon_{\rm L}\left(\omega^{\rm (P)}\right)}d_{\rm NL}/2}e^{i\Delta k_{z,1,1}^{(-,-)}\left(n-\frac{1}{2}\right)d_{\rm TMD}}{\rm sinc}\left[\Delta k_{z;1,1}^{(-,-)}\frac{d_{\rm TMD}}{2}\right]+\\
            &+E_{{\rm NL}}^{-,p}e^{-ik^{\rm (P)}\sqrt{\varepsilon_{\rm L}\left(\omega^{\rm (P)}\right)}d_{\rm NL}/2}e^{-i\Delta k_{z;1,1}^{(+,+)}\left(n-\frac{1}{2}\right)d_{\rm TMD}}{\rm sinc}\left[\Delta k_{z;1,1}^{(+,+)}\frac{d_{\rm TMD}}{2}\right]
            \end{aligned}
            \right\}.
        \end{aligned}
    \end{equation}
\end{widetext}
In particular, 
\begin{equation}
\begin{aligned}
    &\Delta k_{z;f,f'}^{(\pm,\pm)}=k^{\rm (P)}\sqrt{\varepsilon_{\rm L}\left(\omega^{\rm (P)}\right)}+\\
    &\hspace{0.5cm}\pm f\sqrt{\frac{\left(\omega^{\rm (Q)}\right)^2}{c^2}\varepsilon_{\rm L}'-k_{\perp}^2}\pm f'\sqrt{\frac{\left(\omega^{\rm (Q)}\right)^2}{c^2}\varepsilon_{\rm L}'-k_{\perp}'^2},
\end{aligned}
\end{equation}
$\left(\hat{a}_{{\bf k}_{\perp},s}^{\rm VAC}\right)^{\dagger}$ and $\left(\hat{a}_{{\bf k}_{\perp}',s'}^{\rm VAC}\right)^{\dagger}$ are bosonic creation operator of the signal/idler field in the detection region, and $M_s^\sigma$ are refraction matrix elements determined through the transfer matrix method \cite{Yeh1990} such that $\hat{a}_{{\bf k}_{\perp},1,\sigma}=\sum_{s=\pm 1}M_{s}^{\sigma}\hat{a}_{{\bf k}_{\perp},s}^{\rm VAC}$.

We now move to the calculation of the SPDC quantum state. Through a first order Dyson perturbative expansion \cite{Mandel1995}, we are able to determine the biphoton state $|\psi(t)\rangle_N^{\rm tot}=e^{-\frac{i}{\hbar}\hat{H}_{\rm INT}t}|{\rm vac}\rangle\simeq \left(\hat{\mathbb{I}}-\frac{i}{\hbar}\hat{H}_{\rm INT}t\right)|{\rm vac}\rangle=|{\rm vac}\rangle +
|\psi(t)\rangle_N$ ($N$ is the number of slabs) and characterize the polarization properties of the down-converted photons. Finally, to compute the enhancement (ENH) in the down-conversion process, we define
\begin{equation}
{\rm ENH}=\frac{\langle \hat{\rm N}\rangle_{\Sigma,N}}{\langle \hat{\rm N}\rangle_{\Sigma,1}}=\frac{_N\langle \psi(t)|\hat{V}^{\dagger}\hat{V}|\psi(t)\rangle_N}{_1\langle \psi(t)|\hat{V}^{\dagger}\hat{V}|\psi(t)\rangle_1},
\end{equation}
where $\langle \hat{\rm N}\rangle_{\Sigma,N}$ is related to the average number of detected photons in the detection region within a given numerical aperture $\Sigma$ and given $N$ periodically-poled slabs, and
\begin{equation}
    \hat{V}({\bf r},t)=\sum_{s=\pm 1}\int_{\Sigma}\frac{d{\bf k}_{\perp}}{2\pi}\hat{a}_{{\bf k}_{\perp},s}^{\rm VAC}\hat{\bf n}_{{\bf k}_{\perp},s}e^{i{\bf k}_{\perp}\cdot{\bf r}_{\perp}}e^{i\sqrt{\frac{\left(\omega^{\rm (Q)}\right)^2}{c^2}-k_{\perp}^2}},
\end{equation}
where $\hat{\bf n}_{{\bf k}_{\perp},s}$ is the unit polarization vector associated with the specific propagating mode with polarization $s$ in the detection region. Through Eq. (13) we were able to characterize the results shown in Fig. \ref{fig:fig4}(e). 

For what concerns the purity analysis, we can define at first the total un-normalized density matrix as $\hat{\rho}_{\rm tot,U}^{\Sigma,N}=|\psi(t)\rangle_N{_N}\langle \psi (t)|$. We can now retrieve the polarization density matrix by taking the partial trace of $\hat{\rho}_{\rm tot,U}^{\Sigma, N}$ over the spatial degree of freedom, that is 
\begin{equation}
    \begin{aligned}
        &\hat{\rho}_{\rm POL,U}=\sum_{\xi,\xi'=\pm 1}\int_{\Sigma}d{\bf p}_{\perp}\int_{\Sigma}d{\bf p}_{\perp}'\times\\
    &\hspace{1.5cm}\times \langle 1_{{\bf p}_{\perp}',\xi'}^{\rm VAC}|\langle 1_{{\bf p}_{\perp},\xi}^{\rm VAC}|\hat{\rho}_{\rm tot,U}^{\Sigma, N}|1_{{\bf p}_{\perp},\xi}^{\rm VAC}\rangle|1_{{\bf p}_{\perp}',\xi'}^{\rm VAC}\rangle
    \end{aligned}
\end{equation}
where we used $|1_{{\bf k}_{\perp},f,s}^{\rm VAC}\rangle\equiv|1_{{\bf k}_{\perp},f}^{\rm VAC}\rangle|s\rangle$. We can then normalize it and calculate the purity as $P_{\Sigma,N}={\rm tr}\left(\hat{\rho}_{\rm POL}^2\right)$ with $\hat{\rho}_{\rm POL}=\hat{\rho}_{\rm POL,U}/{\rm tr}\left(\hat{\rho}_{\rm POL,U}\right)$.
Eventually, through the comparison of the theoretical model with the experimental results in Fig. \ref{fig:fig4}, we were able to estimate $|\chi^{(2)}_{yzy}|=|\chi^{(2)}_{zyy}|=0.6958|\chi^{(2)}_{xxx}|$.

\subsection{Purity Estimation}\label{app_purity}

We extract the purity data by exciting our sample with vertical polarization, generating the Bell state $\left|\psi^{+}\right>$. Since we have perfect anti-correlation in polarization, the photons will always be split on the PBS. Putting the half-wave plate to the horizontal setting will not affect the polarization of the photons in this case. Rotating the half-wave plate to \ang{22.5}, transforms the $\left|\psi^{+}\right>$ state to $\left|\phi^{-}\right>$. In this case, the photon pairs always bunch on the PBS and no coincidence events between the two detectors will occur. For this idealized case, we therefore expect the following behavior as a function of the wave-plate angle:
\begin{equation*}
    p_{\rm split} = \cos^2(4\alpha),
\end{equation*}
where $p_{\rm split}$ is the probability of the photon pair being routed to different spatial modes on the PBS and $\alpha$ is the wave-plate angle. If we excite the sample with vertically polarized light, we generate the $\left|\phi^{-}\right>$ state which results in the same functional behavior, out of phase relative to the previous case:
\begin{equation*}
    p_{\rm split} = \cos^2\left(4\alpha+\frac{\pi}{2}\right)=\sin^2(4\alpha).
\end{equation*}

If we assume each slab of the PPTMD generates a pure state individually but consequent slabs are not perfectly twisted by \SI{180}{\degree}, we end up with a mixed state $\rho$:
\begin{equation*}
    \rho = (1-\theta)\left|\psi^+\right>\left<\psi^+\right|+\theta\left|\phi^-\right>\left<\phi^-\right|,
\end{equation*}
where $\theta$ is the mixing parameter, which we assume to be small. The purity $P$ of this state $\rho$ is then:
\begin{equation*}
    P(\rho)=\text{Tr}(\rho^2)=1-2\theta(1-\theta).
\end{equation*}
The splitting behavior on the PBS consequently is:
\begin{equation*}
    p_{\rm split}(\rho)=(1-\theta)\cos^2(4\alpha)+\theta\sin^2(4\alpha).
\end{equation*}
From this, we can extract the visibility $V$, which for  $\theta< \frac{1}{2}$ is:
\begin{align*}
    V&=\frac{\text{max}(p_{\rm split})-\text{min}(p_{\rm split})}{\text{max}(p_{\rm split})+\text{min}(p_{\rm split})}=\\
    &=\frac{(1-\theta)-\theta}{(1-\theta)+\theta}=1-2\theta.
\end{align*}
We can now express the purity $P$ of our state $\rho$ as a function of the measured visibility $V$:
\begin{equation}
    P=\frac{1}{2}(1+V^2). \label{eq_purity2}
\end{equation}


\end{document}